\documentstyle[preprint,aps,epsfig]{revtex}

\newcommand{\beq}{\begin{equation}}
\newcommand{\eeq}{\end{equation}}
\newcommand{\beqs}{\begin{eqnarray}}
\newcommand{\eeqs}{\end{eqnarray}}

\def\lsim{\ \rlap{\raise 3pt \hbox{$<$}}{\lower 3pt \hbox{$\sim$}}\ }
\def\gsim{\ \rlap{\raise 3pt \hbox{$>$}}{\lower 3pt \hbox{$\sim$}}\ }

\def\npb#1{Nucl.\ Phys.\ {B\, \bf #1}}

\def\plb#1{Phys.\ Lett.\ {B\,\bf #1}}

\def\prd#1{Phys.\ Rev.\ {D\,\bf #1}}
\def\prl#1{Phys.\ Rev.\ Lett. {\bf#1}}

\def\mpla#1{Mod. Phys. Lett. {A\, \bf #1}}
\def\sjnp#1{Sov. J. Nucl. Phys. {\bf #1}}
\def\JHEP#1{JHEP {\bf #1}}
\def\lsim{\raise0.3ex\hbox{$\;<$\kern-0.75em\raise-1.1ex\hbox{$\sim\;$}}}
\def\gsim{\raise0.3ex\hbox{$\;>$\kern-0.75em\raise-1.1ex\hbox{$\sim\;$}}}

\def\be7{$^{7}$Be}
\def\b8{$^{8}$B}
\begin{document}
\textheight = 23.5cm

\draft
\preprint{\vbox{\hbox{IFUSP-DFN/062}
\hbox{IFT-P.071/2001}
\hbox{hep-ph/0112060}
}}
\title{
Global Analysis of the post-SNO Solar Neutrino Data 
for Standard and Non-Standard Oscillation Mechanisms
}
\author{ 
A.M. Gago$^{1,2,3}$,
M.\ M.\ Guzzo$^{4}$, 
P.\ C.\ de Holanda$^{5}$, 
H.\ Nunokawa$^{4,6}$, 
O.\ L.\ G.\  Peres$^{4}$, \\
V. Pleitez$^{6}$
and
R. Zukanovich Funchal$^{3}$}
\address{\sl ~ \\
$^1$  Secci\'{o}n F\'{\i}sica, Departamento de Ciencias, 
Pontificia Universidad Cat\'olica del Per\'u, \\
Apartado 1761, Lima, Per\'u\\
\vspace{3mm}
$^2$ 
Department of Physics, California State University, 
Dominguez Hills, Carson CA 90747, USA\\
\vspace{3mm}
$^3$ Instituto de F\'\i sica, Universidade de S\~ao Paulo, 
C.\ P.\ 66.318, 05315-970 S\~ao Paulo SP, Brazil \\
\vspace{3mm}
$^4$ Instituto de F\' {\i}sica Gleb Wataghin, 
     Universidade Estadual de Campinas -- UNICAMP \\    
     13083-970 Campinas SP, Brazil \\ 
\vspace{3mm}
$^5$ The Abdus Salam International Center for Theoretical Physics, 
     I-34100 Trieste,  Italy \\ 
\vspace{3mm}
$^6$ Instituto de F\'\i sica Te\'orica, 
     Universidade Estadual Paulista \\
     Rua Pamplona 145,      01405-900 S\~ao Paulo SP, Brazil}

\maketitle
\vspace{.5cm}
            
\hfuzz=25pt

\begin{abstract}

\noindent
What can we learn from solar neutrino observations? 
Is there any solution to the solar neutrino anomaly which 
is favored by the present experimental panorama? 
After SNO results, is it possible to affirm that
neutrinos have mass? 
In order to answer such questions we analyze the current available data 
from the solar neutrino experiments, including the
recent SNO result, in view of many
acceptable solutions to the solar neutrino problem based on different
conversion mechanisms, for the first time,  
using the same statistical procedure. This allows 
us to do a direct comparison of the goodness 
of the fit among different solutions, from which 
we can discuss and conclude on the current status 
of each proposed dynamical mechanism.
These solutions are based on different assumptions: 
(a) neutrino mass and mixing, 
(b) non-vanishing neutrino magnetic moment, 
(c) the existence of non-standard flavor-changing 
and non-universal neutrino interactions and 
(d) the tiny violation of the equivalence principle. 
We investigate the quality of the fit provided by 
each one of these solutions not only to the total rate 
measured by all the solar neutrino experiments 
but also to the recoil electron energy spectrum 
measured at different zenith angles by 
the Super-Kamiokande collaboration. 
We conclude that several non-standard neutrino flavor 
conversion mechanisms provide a very good fit to the experimental 
data which is comparable with (or even slightly better 
than) the most famous solution to the solar neutrino 
anomaly based on the neutrino oscillation induced by mass.

\end{abstract} 
\newpage

\section{Introduction}
\label{sec:intro}

Solar neutrino observations coming from Homestake~\cite{Lande:nv}, 
Kamiokande~\cite{Fukuda:1996sz}, SAGE~\cite{Abdurashitov:uu}, 
GALLEX~\cite{Hampel:1998xg}, GNO~\cite{Altmann:2000ft}, 
and Super-Kamiokande (SK)~\cite{Fukuda:1998rq} have been suggesting 
a picture which is conflicting with the predictions from 
the Standard Solar Model (SSM)~\cite{JNB,BP00,concha-novo}, 
strongly indicating disappearance of solar electron neutrinos 
on their way from the Sun to the terrestrial detectors. 
This has been known for many years as the {\em Solar Neutrino 
Problem} (SNP)~\cite{JNB}.

The extraordinary new result from the Sudbury Neutrino Observatory 
(SNO)~\cite{SNO}, inaugurates a new era in the quest for the solution to 
the long-standing puzzle of missing solar neutrinos.
For the first time in the history of solar neutrino observations,
a direct indication of the presence of non-electron active 
neutrino component in the solar neutrino flux is obtained.
This cannot be explained by any conceivable modification 
of the SSM but do require some 
departure from the standard electroweak theory. 
The indication of non-electron active component is 
based on the difference of $^8$B neutrino flux 
detected through charged current events in SNO and 
the neutrino electron elastic scattering events 
observed by the SK collaboration, the former 
obtaining a lower rate than the latter. 
Such difference can be explained by the conversion of
electron neutrinos into active non-electron 
($\nu_\mu$ or $\nu_\tau$) neutrinos along 
their trajectory from the Sun to the detectors at the 
Earth\cite{SNO,fogli_sno}.

As we will see, the hypothesis of no flavor conversion of 
solar electron neutrinos is strongly in conflict with 
the prediction of the SSM; it is now only acceptable at very 
small confidence level ($\sim 10^{-12}$) if only the 
total rates from solar neutrino experiments is considered 
(see Sec.\ \ref{sec:resul}). 
The disagreement among the observed  solar neutrino data 
and the theoretical predictions can be relaxed to 
4~$\sigma$ level ($7 \times 10^{-5}$)~\cite{Bahcall2001}  
if one allows all the solar neutrino fluxes
to be free parameters in fitting 
the measured solar neutrino event rates.
However, this can only be obtained under the extreme assumption  
of vanishing  $^{7}$Be neutrino flux, which is quite difficult 
to explain.

Several mechanisms can induce neutrino flavor
conversion when one assumes that neutrinos are 
endowed with some properties not present 
in the minimal  standard electroweak theory~\cite{30years}.  
The most well known mechanism is the neutrino oscillation 
induced by mass and mixing~\cite{MNS,vacuum,wolf,MS}. 
The fact that the terrestrial experiments are less sensitive 
to these resulting non-electron neutrinos can explain 
their observed lower counting rates.   
The purpose of this article is to compare quantitatively the
capabilities of several different mechanisms to explain solar neutrino
data, for the first time,  based on the same 
statistical procedure. 

Readers are invited to take a look at Tables \ref{tab:rates} 
and \ref{tab:combined} which summarize our most important results. 
A number of possible solutions to the solar neutrino anomaly still 
survives even after SNO results, fitting the data with 
significant confidence level. 
In fact, combined analysis of the data, which includes not only 
the total rates  measured by all solar neutrino experiments but 
also some information which is independent of the total neutrino flux, 
namely,  the energy spectrum and the zenithal dependence of the data, 
suggests that the large mixing angle MSW solution in matter, 
as well as the mechanisms based on resonant spin-flavor precession, 
non-standard neutrino interactions and 
violation of the equivalence principle all provide a fit of 
the data with the confidence level $\gsim$ 60\%.  

In Sec.\ \ref{sec:mec}, we briefly review several mechanisms that 
induce flavor conversion of solar neutrinos, which will be
discussed in this work: (a) mass induced oscillation in vacuum and 
in matter (b) resonant spin-flavor precessions  
induced by a non-vanishing neutrino magnetic moment, 
(c) the existence of  non-standard neutrino interactions  
inducing flavor-changing and non-universal currents and 
(d)  the violation of the equivalence principle. 
In Sec.\ \ref{sec:proc}, the procedure of the statistical analysis 
used in this work is presented, while our results are given in 
Sec.\ \ref{sec:resul}. Finally, Sec.\  \ref{sec:conc} is devoted to 
the discussion of our results.

\section{Neutrino Conversion Mechanisms}
\label{sec:mec}

The neutrino conversion mechanisms we will consider in 
this work can be phenomenologically described 
by the following Schr\"oedinger-like evolution equation: 
\begin{equation} i \frac{d}{dr}
\left[
\begin{array}{l}
\nu_{e} \\
\tilde{\nu}
\end{array}
\right]
=
\left[
\begin{array}{cc}
A(r) \,  \,   &   C(r)    \\
C(r)     \,  \,   &   -A(r)
\end{array}
\right]
\left[
\begin{array}{l}
\nu_{e}   \\
\tilde{\nu} 
\end{array}
\right]  ,
\label{evolution}
\end{equation}
where $r$ is the distance travelled by neutrino, 
$\nu_e$ is the initial electron neutrino state, 
$\tilde\nu$ is the neutrino state to which the conversion 
mechanism leads and the explicit form of the elements 
of the $2\times 2$ Hamiltonian matrix, $A(r)$ and $C(r)$, 
which are in general depend on the position $r$, 
will be given in the next subsections for each 
conversion mechanism. 
For simplicity, we assume neutrino conversion only between 
two flavors, or more precisely, we assume that neutrino 
conversion relevant for solar neutrinos can be effectively 
described, in good approximation, in terms
of two flavor conversion.

In all the mechanisms, either in vacuum or in matter, 
the phenomenon of neutrino mixing occurs and this  can be generally 
expressed as 
\begin{equation} 
\left[
\begin{array}{l}
\nu_{e} \\
\tilde{\nu}
\end{array}
\right]
=
\left[
\begin{array}{cc}
\cos\tilde{\theta} \,  \,   & \sin\tilde{\theta}    \\
-\sin\tilde{\theta}  \,  \,   &   \cos\tilde{\theta}
\end{array}
\right]
\left[
\begin{array}{l}
\nu_{1}   \\
\nu_{2} 
\end{array}
\right]  ,
\label{mixing}
\end{equation}
where $\tilde{\theta}$ is the mixing angle which relate
neutrino flavor and the propagating eigenstates $\nu_1$ and $\nu_2$  in 
matter
or in vacuum. Note that $\tilde{\theta}$ can be defined 
even for massless neutrinos as in the case of NSNI or VEP. 

For all the mechanisms we analyze in this work, we do not 
consider the case where $\tilde{\nu}$  is a sterile 
(or electroweak singlet) neutrino~\cite{sterile}   
because of the present indication of the presence of non-electron active 
component in the solar neutrino flux, which is provided by the combined
information from SNO and SK results. 
Namely, neutrino conversion of solar $\nu_e$ only 
into a sterile neutrino is not favored by 
the current data. 

\subsection{Mass induced oscillation (MIO)}
\label{subsec:mio}

Let us first consider the most popular conversion mechanism, 
the one induced by neutrino mass and mixing. 
If one assumes that neutrinos are massive, 
the flavor eigenstates do not coincide, 
in general, with the mass eigenstates, leading to neutrino flavor 
oscillation in vacuum~\cite{MNS,vacuum} as well as in matter where
it can be resonantly enhanced~\cite{wolf,MS}. The latter is 
known as the Mikheyev-Smirnov-Wolfenstein (MSW) effect. 
It is generally believed that this is the most plausible 
mechanism which can induce neutrino oscillation since 
to introduce mass and mixing in the leptonic sector 
is the simplest and most natural extensions to the 
standard electroweak model, so theoretically well motivated. 

In both scenarios, MIO can convert solar electron 
neutrinos into neutrinos of a different flavor and 
consequently explain the deficit of observed solar neutrino 
with respect to the predictions of the SSM. 
We will refer to these type of solutions  as standard solutions 
to the SNP,  they are very well described 
in recent Refs.~\cite{BKS,GNZ3gvac}. 
MIO requires in Eq.~(\ref{evolution}) that
\begin{equation}
\begin{array}{l}
A(r)= - \displaystyle{\delta m^2 \over 4E} 
\cos 2 \theta + {G_F \over \sqrt{2}}N_e(r),  
\\ C(r)= \displaystyle{\delta m^2 \over 4E}\sin 2\theta,
\end{array}
\end{equation}
where $\delta m^2 \equiv m^2_2-m^2_1$ is the mass squared 
difference of the two neutrinos involved, $\theta$ is the
vacuum mixing angle, $G_F$ is the Fermi constant,  
$E$ the neutrino energy  and $N_e(r)$ is the electron number 
density at position $r$, 
and here $\tilde \nu$  is identified with $\nu_\mu$ or $\nu_\tau$ 
(or their linear combination).
We compute the conversion probability using the analytic formulas, 
properly taking  into account the neutrino production distributions, 
as well as the Earth matter effect as in Ref.~\cite{GHPV}.
The relevant oscillation parameters, which must be 
determined by the fit to experimental data, 
are $\delta m^2$ and  $\theta$. 
Recent post-SNO analyses can be found in Ref.~\cite{analysis_sno}. 
Depending on the allowed parameter values, 
the solutions based on this mechanism are classified into 
large mixing angle (LMA) MSW, small mixing angle (SMA) MSW, 
low-$\delta m^2$ (LOW) MSW and vacuum oscillation (VAC) solutions.

\subsection{Resonant spin-flavor precession (RSFP)}
\label{subsec:rsfp}

Assuming neutrinos have a non-vanishing transition magnetic moment,
electron neutrinos interacting with the solar magnetic field 
can be spin-flavor converted into active non-electron 
anti-neutrinos~\cite{schechtervalle}, if they are of Majorana type. 
Here we do not consider the case of Dirac type neutrinos 
since it involves a sterile one. 
Furthermore, such spin-flavor precession of neutrinos can be 
resonantly enhanced in matter~\cite{LimMarciano}, 
in close analogy to the MSW effect~\cite{wolf,MS}. 
RSFP could strongly depend on the neutrino energy and 
provoke different suppressions for each portion of the 
solar neutrino energy spectrum, in a similar way as in the 
case of the MSW effect. 
In fact it has been known that RSFP could provides a satisfactory 
description~\cite{review,rsfp_analyses,GN98} of 
the actual experimental 
panorama~\cite{Lande:nv,Fukuda:1996sz,Abdurashitov:uu,Hampel:1998xg,Altmann:2000ft,Fukuda:1998rq}.
However, see a recent work~\cite{miranda} where it is 
found that the non-resonant spin-flavor precession can also explain
the solar neutrino data. 

The time evolution of neutrinos interacting with a magnetic field $B$ 
through a non vanishing neutrino magnetic moment $\mu_\nu$ in matter 
is governed by Eq.~(\ref{evolution}) by 
defining $A(r)$ and $C(r)$ as \cite{LimMarciano}:

\begin{equation}
\begin{array}{l}
A(r)= -\displaystyle \frac{\delta m^2}{4E} 
+ \frac{1}{\sqrt{2}}G_F\left[ N_e(r)-N_n(r)\right], \\
C(r)=\displaystyle \mu_\nu B(r),    
\end{array}
\end{equation}
where $N_n(r)$ is the neutron number densities, 
$\mu_\nu$ is the transition magnetic moment between two neutrino
involved, $B(r)$ is the solar magnetic field and 
here $\tilde \nu$  is identified with $\bar{\nu}_\mu$ 
or $\bar{\nu}_\tau$ (or their combination). 
We assume, for simplicity that the mixing angle is zero in this
mechanism. 
Note that, roughly speaking, $\mu_\nu B(r)$ is playing 
the same role as the mixing term 
$({\delta m^2/4E})\sin 2\theta$,
which appeared in the standard mass induced oscillation
mechanism. 
In this work, we assume, as a reference value, 
the magnitude of the neutrino magnetic moment to be 
$\mu_\nu = 10^{-11} \mu_B$ ($\mu_B$ is the Bohr magneton). 
Since the relevant quantity is only the product 
of the neutrino magnetic moment and the magnetic field, 
it must be understood that if $\mu_\nu$ is taken to be smaller, 
the solar magnetic field must be properly increased to achieve the 
same effect. 

The RSFP mechanism crucially depends on the solar magnetic field profile
along the neutrino trajectory. 
In our present analysis we assume a particular typpe of solar magnetic 
field profile which could explain well the solar neutrino data 
(before SNO result) with relatively weak magnetic field 
($\sim O(10)$ kG), which was considered in number of 
works in Refs.~\cite{rsfp_analyses,GN98}. 
The profile we will use has a triangular configuration 
in the solar convective zone. 
For definiteness, we take the one which is found in 
Ref.~\cite{PK00}. 
The profile has a vanishing magnetic field in 
the internal part of the Sun, linearly growing from $r=0.7$ to
$r=0.85$ where it achieves its maximum value $B_{\text{max}}$ 
and begins to linearly  decline  until  
the surface of the Sun, where $r=1$. Here $r$ is the radial distance from 
the center of the Sun normalized by the solar radius. 
In this work, we consider $B_{\text{max}}$ up to 500 kG taking
into consideration the upper limit (300 kG) of the magnetic field 
at the bottom of the convective zone obtained 
in Ref.~\cite{Antia00}. 
Once the shape of the magnetic field profile is fixed, 
the relevant parameters which must be determined by 
the fit can be chosen to be  $\delta m^2$ and $B_{\text{max}}$. 

We compute the conversion probability by numerically integrating
the evolution equation as in Ref.~\cite{GN98}. For simplicity, 
we assumed that all the neutrinos are created in the solar center 
and neglected the Earth matter effect, as discussed in Ref.~\cite{GN98}.

\subsection{Non-standard neutrino interactions (NSNI)}
\label{subsec:nsni}

In his seminal paper Wolfenstein~\cite{wolf} observed that 
NSNI with matter can also generate neutrino oscillation
even without flavor mixing in vacuum. 
Some explicit examples of such NSNI induced neutrino oscillation
were considered in Refs.~\cite{GMP,BPW}. 
It has been shown that NSNI could be relevant to solar neutrinos
propagating in the solar matter along their path from 
the core of the Sun to its 
surface~\cite{GMP,BPW,BGHKN}
as well as in the Earth until they reach the detector. 

The evolution equations for massless neutrinos 
(or neutrinos with degenerate mass) 
having such NSNI in matter can be phenomenologically 
expressed by Eq.~(\ref{evolution}) with the following 
definitions of $A(r)$ and $C(r)$~\cite{GMP,BPW}:
\begin{equation}
\begin{array}{l}
A(r)=  \displaystyle \frac{1}{\sqrt2} 
G_F [N_e(r)- {\epsilon'}_{f} N_f(r)], \\
C(r)=\sqrt2 G_F \epsilon_{f} N_f(r),  
\end{array}
\end{equation}
where ${\epsilon'}_{f}$ and ${\epsilon}_{f}$ are 
the phenomenological parameters which characterize 
the strength of the NSNI with fermion $f$ whose number
density is given by $N_f(r)$ with $f =u$ or $d$ quark, 
and here $\tilde \nu$ is identified with 
$\nu_\mu$ or $\nu_\tau$  (or their combination).

Here, by taking into account the charge neutrality, 
$N_f$ can be written in terms of electron and neutron 
number densities, as follows, 
\beq
N_f(r)=\left\{\matrix{N_n(r) + 2N_e(r) & (f=u) \cr
                      2N_n(r) + N_e(r) & (f=d).
} \right. 
\eeq 

The parameters ${\epsilon}_{f}$ appearing in the off-diagonal elements 
of the Hamiltonian matrix is responsible for flavor changing neutrino 
interactions, 
it plays a similar role to the mixing term 
$({\delta m^2/4E})\sin 2\theta$ in the MIO mechanism. It exists even 
if there is no neutrino mixing in vacuum. 
On the other hand, ${\epsilon'}_{f}$ appearing in the 
diagonal element which is responsible for 
flavor diagonal neutrino interactions 
with matter~\cite{GMP}, somehow plays 
a role similar to the mass squared difference in the 
MIO mechanism,  since this term leads to resonantly enhanced conversion 
when its magnitude coincides with that of 
the standard electroweak neutrino interactions
at some point $r_{\text{res}}$ along the neutrino trajectory, 
satisfying the resonance condition, 
\beq 
{\epsilon'}_f N_f(r_{\text{res}}) = N_e(r_{\text{res}}) \,.  
\label{res}
\eeq
An immediate consequence is that if $f$ was to be identified 
with electron, it cannot induce resonant neutrino flavor 
conversion\cite{FC_eletrons}. 

An important characteristic of this mechanism is that the 
conversion probability 
does not depend on neutrino energy as it is understood
from the evolution equation. 
Nevertheless this mechanism can explain quite well 
the solar neutrino data which imply strong energy 
dependent conversion. 
The reason is that even though the conversion probability 
itself is completely energy independent, 
after taking average of the probability over 
production distributions, different neutrinos from 
different nuclear reactions in the Sun ($pp$, $^8$B, $^7$Be, etc.)
can have different final average probability due to the fact that 
their production distributions are different, 
provided that resonant conversion occur close to 
the solar core ($\sim 10 \%$ or so of the solar radius).

Following Ref.~\cite{BGHKN}, we first compute the conversion
probability using the analytic formulas for a given production
point in the Sun and then take average over the production
distribution for each neutrino source. We also take into account
the Earth regeneration effect, which is important for some region
of the parameter space.

As it was observed in Ref.~\cite{GMP}, only $\nu_e \to \nu_\tau$
conversion are compatible with the existing phenomenological
constraints on $\epsilon_f$ and ${\epsilon'}_f$. In this way, the
relevant parameters which must be determined by the fit
for this mechanism are ${\epsilon'}_{f}$ and ${\epsilon}_{f}$.

\subsection{Violation of the equivalence principle (VEP)}
\label{subsec:vep}

It has been proposed that violation of the equivalence principle 
could induce neutrino flavor oscillation even if neutrinos 
are massless \cite{Gasperini,hl}. 
In this scenario, neutrino mixing and flavor oscillation 
can be induced if two (or more) neutrinos involved have
different gravitational couplings which imply VEP. 
In this case, weak interacting eigenstates and gravitational 
interacting eigenstates can be different and will 
be related by a unitary transformation that can be parameterized, 
assuming only two neutrino flavors, by a single parameter, the mixing 
angle $\theta_G$ similar to the case of neutrino mixing in vacuum 
induced by mass.

The evolution equations for these flavors, 
which are assumed to be degenerate in mass, propagating through the
gravitational potential $\phi(r)$ in the absence of matter 
is given by Eq.~(\ref{evolution}) with~\cite{Gasperini,hl}:

\begin{equation}
\begin{array}{l}
A(r) = 2E\phi(r) \delta\gamma \cos 2\theta_G,
\\ C(r)= 2E\phi(r) \delta\gamma \sin 2\theta_G,
\end{array}
\end{equation}
where $\delta\gamma$ is the quantity which measures 
the magnitude of the violation of the equivalence 
principle, the difference of the gravitational 
couplings between the two neutrinos involved 
normalized by the sum.  Here $2E\phi(r) \delta\gamma \sin 2\theta_G$
plays a similar role to the mixing term 
$({\delta m^2/4E})\sin 2\theta$ in the MIO mechanism, this means 
the VEP mechanism gives rise to an oscillation length inversely 
proportional to the neutrino energy $E$, while MIO expects it to 
be directly proportional to $E$. 

As in Ref.~\cite{GNZvep}, we take $\phi(r)$ to be constant
($\sim 10^{-5}$), 
assuming that the local supercluster contribution
is the dominant one~\cite{Kenyon}.
The relevant parameters which must be determined by the fit
in this mechanism are $|\phi \delta\gamma|$ and $\theta_G$.
We consider the product of $\phi$ and $\delta\gamma$ since
the former has a large uncertainty and only the product 
is relevant in the fit. 
Similar to the mass induced oscillation, we can have two types for the
VEP mechanism: (a) VEP induced MSW-like resonant conversion and 
(b) VEP vacuum conversion. 
The former was first discussed in Ref.~\cite{hl}, and then 
analyzed by several authors~\cite{vepmsw}. 
However, Ref.~\cite{PKM} showed that it is not a good solution 
because the required magnitude of the parameters is
incompatible with the CCFR experiment results~\cite{ccfr} 
which exclude $| \phi \delta\gamma|$ larger than  $\sim 10^{-23}$.
For the latter one, it has been shown that this mechanism is a
new solution for the solar neutrino anomaly~\cite{GNZvep}. 
Recent analysis of VEP vacuum solution was discussed 
in Ref.~\cite{vep_recent}.

We calculate the conversion probability using the analytic formulas
as in Ref.~\cite{GNZvep}. Similar to the case of vacuum oscillation solution
in the MIO scenario, we can neglect, as good approximations,
the neutrino production distributions and the Earth matter effect~\cite{GNZvep}.

\section{Statistical Procedure}
\label{sec:proc}

Our main goal is to determine the allowed values of 
the relevant parameters for each one of the mechanisms 
that can explain the experimental observations 
without modifying the SSM predictions and 
to evaluate the quality of the fit. 
In order to achieve our goal, we adopt the minimum $\chi^2$ 
statistical treatment of the data following the description 
found in Ref.~\cite{FL95} which was also employed 
in Refs.~\cite{GHPV,GNZ3gvac,GN98,BGHKN,GNZvep,pedro_phd} but with 
some modifications. 

In this work, we perform three kinds of analysis using 
three $\chi^2$ functions, 
$\chi_R^2$ for the analysis of the rates only, 
$\chi_{\text{fi}}^2$ for the flux independent analysis, 
and  $\chi_{\text{comb}}^2 \equiv  \chi_R^2 + \chi_{\text{fi}}^2$ for 
the combined analysis, where the definitions of 
$\chi_R^2$ and $\chi_{\text{fi}}^2$ are given in the next subsections. 
For each case, we minimize the $\chi^2$ function
in order to determine the best fitted parameters as 
well as its minimum value $\chi_{\text{min}}^2$, which is
relevant for the evaluation of the goodness of fit. 
The allowed parameter region can be determined by the
condition,  
\beq
\chi^2 < \chi^2_{\text{min}} + \Delta \chi^2 \,\,,
\eeq
where $\Delta \chi^2 = 4.61$, $5.99$ and $9.21$ 
for $90\%$, $95\%$ and $99\%$ C.L., respectively.

\subsection{$\chi^2$ for the analysis of the rates}
\label{subsec:rates}

First we describe the $\chi^2$ for the analysis of 
the total event rate measured by the Chlorine (Cl) experiment~\cite{Lande:nv}, 
the Gallium  detectors GALLEX/GNO~\cite{Hampel:1998xg,Altmann:2000ft} 
(we use the combined results of GALLEX and GNO) 
and SAGE~\cite{Abdurashitov:uu}, the water Cherenkov experiment 
SK~\cite{Fukuda:1998rq}  and also the recent neutrino-deuterium charged
current data 
by SNO~\cite{SNO}. 
For simplicity we do not consider the result from the Kamiokande 
experiment~\cite{Fukuda:1996sz} as it is consistent with the current SK one 
and has much larger experimental errors. 
All the solar neutrino data for the analysis of the rates used  
in this work are summarized in Table \ref{tab1}. 
In all our calculation of the theoretical predictions we have used
 the latest ``BP2000 SSM+New $^8$B''~\cite{BP00,concha-novo} fluxes, 
which include the recent the new measurement of $S_{17}(0)$~\cite{junghans}. 
Our $\chi^2$ function is defined as follows:
\begin{equation}
 \label{chi}
\chi^2_R = \sum_{i,j=1,...,5}
 \left[R_i^{\text{th}}-R_i^{\text{obs}} \right] \, 
 \left[\sigma^2_R \right]^{-1}_{ij} \,
 \left[R_j^{\text{th}}-R_j^{\text{obs}} \right]\,,
\end{equation}
where $R_i^{\text{th}}$ and $R_i^{\text{obs}}$ denote, respectively, 
the predicted and the measured value for the event rates of the five solar
experiments considered, $i$ = Cl, GALLEX/GNO, SAGE, SK, SNO. 

In order to compute the predictions for the  rates, for Chlorine,
Gallium and SK detectors, we follow Refs.~\cite{GHPV,GNZ3gvac}. 
For SNO, we compute the rates using the neutrino charged current 
cross section on deuterium, $\nu_e + d \to p + p + e^-$, as given in 
Ref.~\cite{nue_xsection} and taking into account the energy resolution as 
described in Ref.~\cite{SNO}.
The error matrix $\sigma_R$ contains both experimental 
(systematic and statistical) and theoretical errors~\cite{FL95}.

\subsection{$\chi^2$ for flux independent analysis: SK zenith and spectrum}
\label{subsec:spec}

Our final result is the fit derived from the combined analysis of 
all presently available solar neutrino data, which includes the flux
independent information presented by SK Collaboration ~\cite{sk_zensp}. 
We include this information using the SK data split simultaneously 
on 7 zenith bins and 8 spectrum bins. We will refer to these as 
the SK zenith-spectrum data. 
These data are summarized in Table \ref{tab:spzenith}. 
For the first and last spectrum bins
the zenith data are grouped together in one data point, due to
low detection rate at these bins, in total 44 data points related
to flux independent information. We took the statistical and
systematic energy-correlated and uncorrelated uncertainties, listed
in Ref.~\cite{sk_zensp}. 
The systematic uncertainties are assumed to be fully
correlated in the zenith angle splitting. We then use the following
$\chi^2$ expression for the flux-independent information:
\begin{equation}
\chi^2_{\text{fi}}=\sum_{i=1,44} 
[\alpha R^{\text{th}}_{\text{SK},i}-R^{\text{obs}}_{\text{SK},i}]
[\sigma_{\text{fi}}(i,j)^2]^{-1}
[\alpha R^{\text{th}}_{\text{SK},i}-R^{\text{obs}}_{\text{SK},i}]
\,\,,
\label{eq:chi2spzen}
\end{equation}
where $R^{\text{th}}_{\text{SK},i}$ is the theoretically 
expected event rates for the $i$-th bin computed by 
using the $^8$B neutrino energy spectrum 
given in Ref.~\cite{NewBspec} normalized to  
the BP2000 SSM value~\cite{BP00}, 
$R^{\text{obs}}_{\text{SK},i}$ is the corresponding observed rate reported
by the SK Collaboration~\cite{sk_zensp}, 
$\alpha $ is a free parameter to avoid double-counting 
of the SK total rate in the statistical treatment, 
and $\sigma_{\text{fi}}(i,j)$ is the 44 $\times$ 44 
error matrix for SK zenith-spectrum data. 
This treatment is slightly different from the one presented in recent
solar neutrino analysis~\cite{concha-novo,fogli_sno,analysis_sno} 
which used the SK elastic scattering spectrum data taken during 
the day and the night separately
whereas we used the SK spectrum measured at different 
zenith angle (night spectrum data are divided into 6 zenith bins). 

\section{Results}
\label{sec:resul}

In this section, we discuss the results of our statistical analysis. 
In Table \ref{tab:rates}, a comparison of the existing solutions to 
the SNP, when only the total rates are taken 
into consideration in the statistical analysis is presented. 
For each one of the indicated mechanisms the best fit values 
of the relevant parameters  are shown, followed by the
corresponding $\chi^2_{\text{min}}$ and its confidence level. 
In the first row one can find the result of the poor fit of the SSM
predictions to the data~:
$\chi^2_{\text{min}}$ = 62.5 for 5 d.o.f. which represents a confidence 
level of only $\sim 10^{-12}$. 
All the other solutions present 2 free parameters to
fit 5 experimental data points, resulting in 3 d.o.f.. 
The largest confidence level (87\%), showed in Table \ref{tab:rates},
of the fit to the observed data is achieved by the solutions 
based on neutrino NSNI and the second largest one (80\%) 
is achieved by the RSFP solution. 
We note that the RSFP solution has an extra freedom in choosing 
the solar magnetic field profile.  
In deed we found that the RSFP solution is rather sensitive to 
the relatively small change of the magnetic field profile. 
We found that if we use the similar but different 
magnetic field profile used in Ref.~\cite{GN98}, which has 
the peak at $r=0.8$, somewhat inner region of the sun, 
solutions exist only for $B_{\text{max}}$ larger than $\sim 100$ kG.
Among the solutions with no such extra freedom, MIO in vacuum
is the second best one (45 \%).
     
For the no-oscillation hypotheses, in the case where only 
the SK zenith-spectrum information is used, 
we obtain $\chi^2_{\text{fi}}=37.7$ 
with $\alpha =0.446$, for $43$ d.o.f., which is compatible 
with the experimental data at $70\%$ C.L..
This results imply that there are no strong distortions 
in the the shape of SK zenith-spectra when compared to  
the SSM predictions, and the observed
data are consistent with no-oscillation hypothesis 
provided that the overall rate is normalized. 

When we include in the statistical analysis not only the total rates
but also the energy spectrum and 
zenithal dependence of the data coming from the SK experiment, then the
panorama changes as summarized in 
Table~\ref{tab:combined}. As in Table~\ref{tab:rates}, the first row here shows 
the fit of the SSM predictions to the 
combined data: 
$\chi^2_{\text{min}}$~= 100.0 for 48 d.o.f., which represents 
a confidence level of $\sim 10^{-5}$. 
The solutions listed in Table~\ref{tab:combined} 
present 2 free parameters to fit 48 experimental data points, 
resulting in 46 d.o.f.. 
Here, the large mixing angle MSW solution based on MIO, 
as well as RSFP and NSNI solutions are equally
probable all having  $\gsim$ 80\% C.L.. VEP solution also 
provides a very good fit, $60$ \% C.L. 

In Figs. 1(a)-5(a), we present, for all the conversion mechanisms, 
(i) the allowed parameter region determined 
from the total rates only and 
(ii) the excluded parameter region determined only 
from the SK zenith-spectrum information. 
In Figs. 1(b)-5(b), we present the allowed parameter region 
determined from the combined data of the rates and 
the SK zenith-spectrum information. 

\begin{figure}[ht]
\begin{center}
\epsfig{file=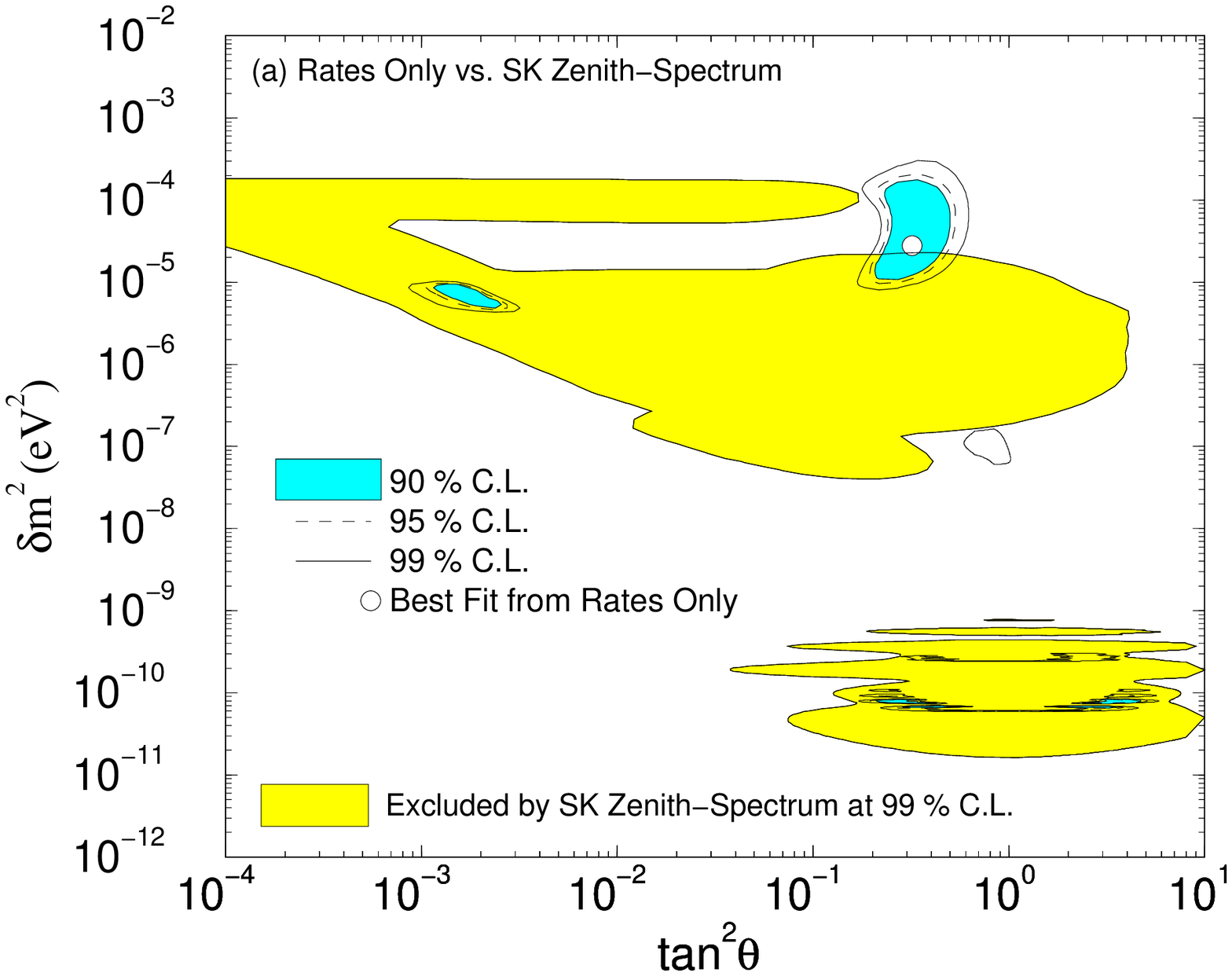,height=11.0cm,width=11.0cm}  
\vglue -1.00cm  
\epsfig{file=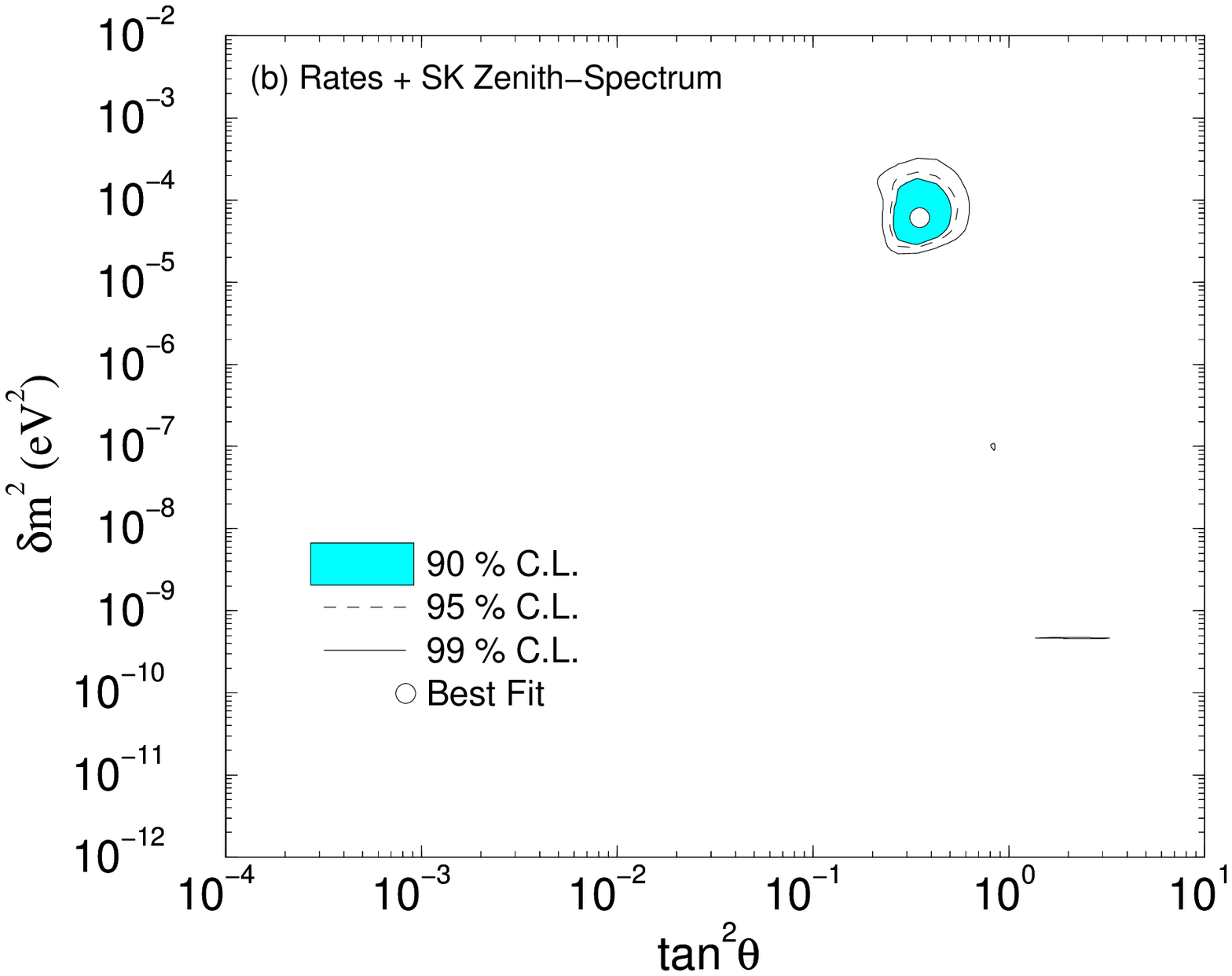,height=11.0cm,width=11.0cm}  
\end{center}
\vglue -0.6cm  
\caption{ 
In (a) the allowed region for the MIO solutions to the SNP 
at 90, 95 and 99 \% C.L. 
with rates only as well as the excluded region at 99 \% C.L. from 
the SK zenith-spectrum information are shown. 
In (b) the allowed region from the combined information from 
rates and zenith-spectrum data is presented. 
} 
\label{fig:mio} 
\end{figure}

\begin{figure}[ht] 
\begin{center}
\epsfig{file=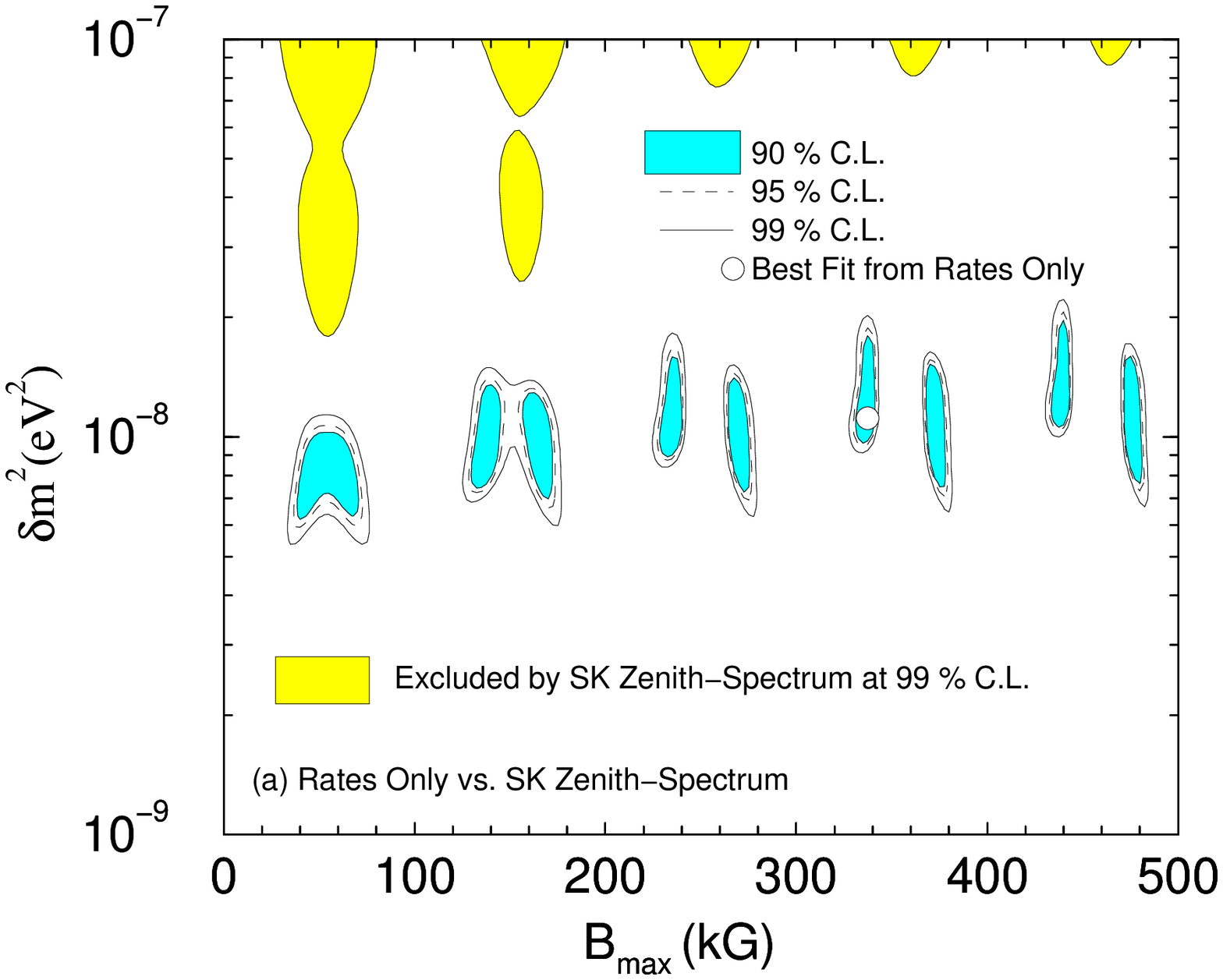,height=10.0cm,width=11.0cm}  
\vglue 0.2cm  
\epsfig{file=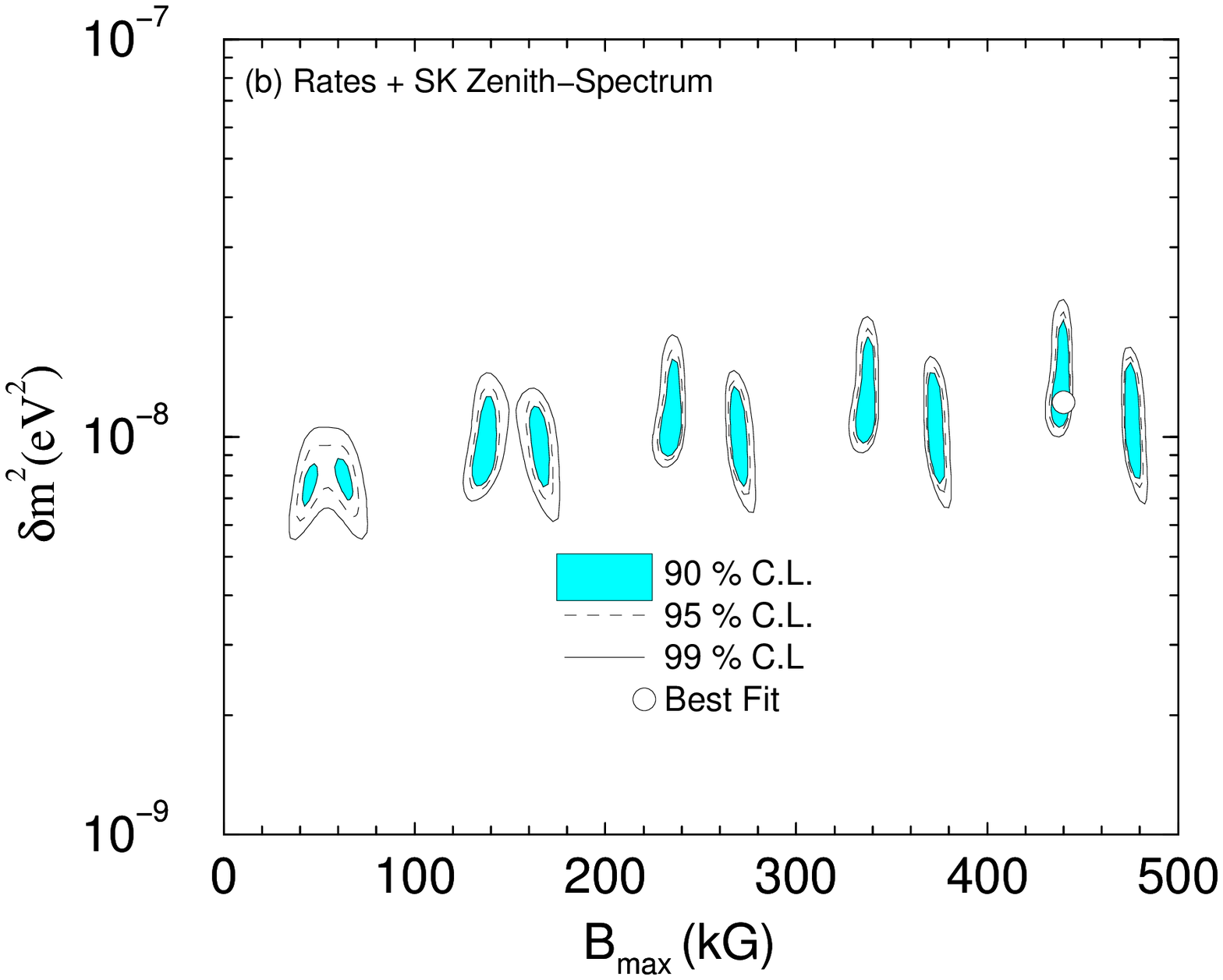,height=10.0cm,width=11.0cm}  
\end{center}
\vglue -0.2cm  
\caption{ 
Same as Fig.~\ref{fig:mio} but for the  RSFP solution
in the $B_{max}-\delta m^2$ plane.} 
\label{fig:rsfp} 
\end{figure}

\begin{figure}[ht] 
\begin{center} 
\epsfig{file=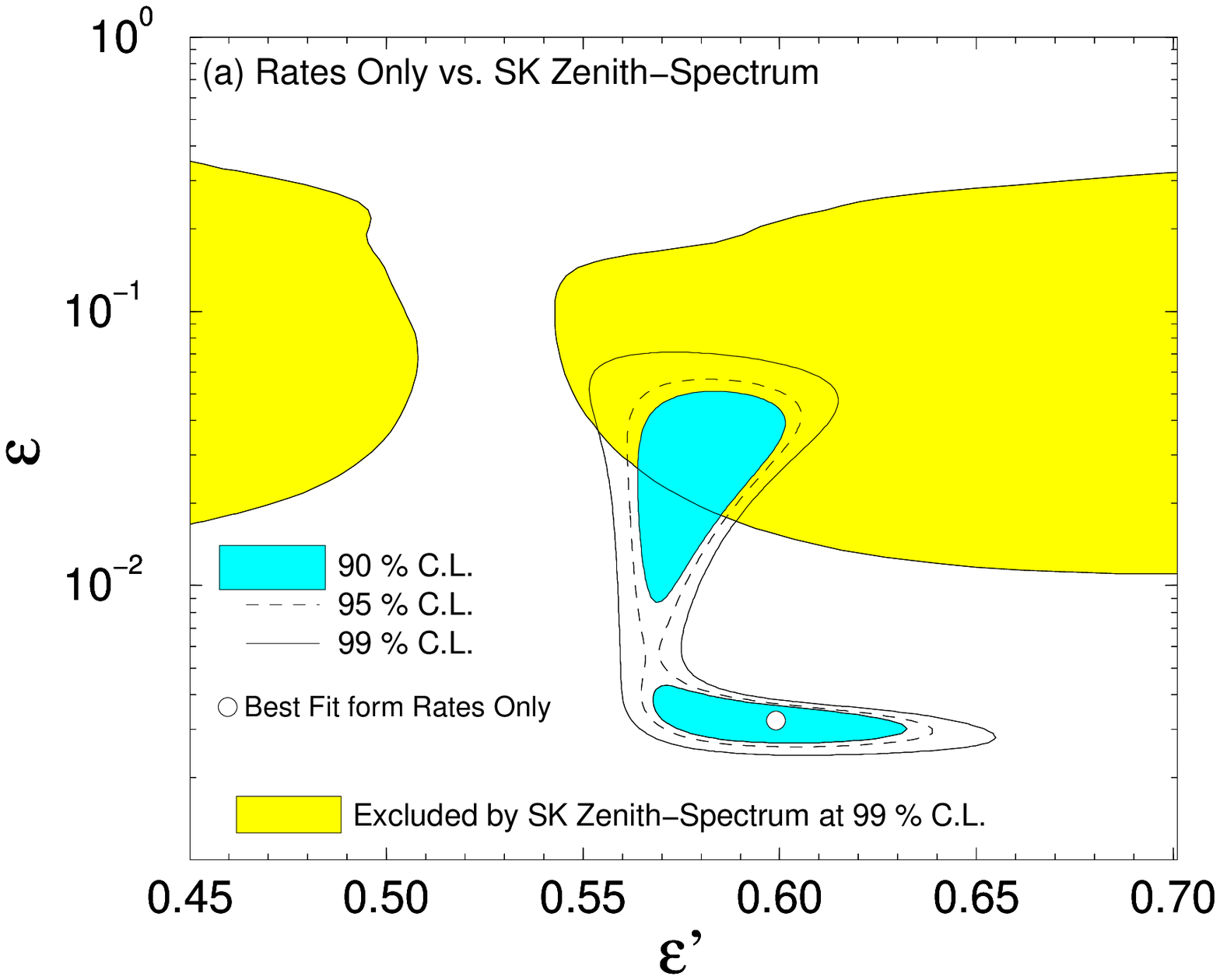,height=10.0cm,width=10.0cm}  
\epsfig{file=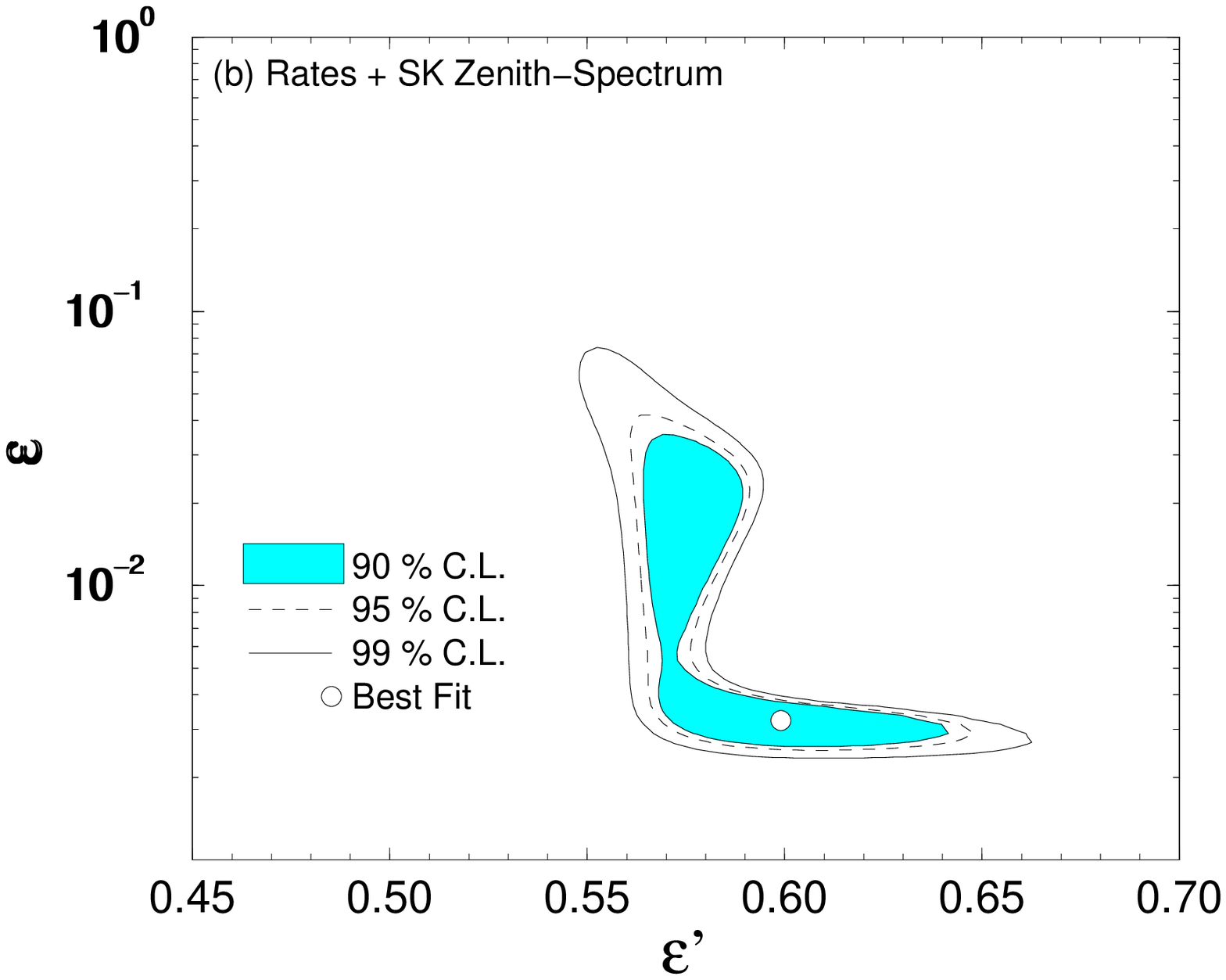,height=10.0cm,width=10.0cm}  
\end{center} 
\vglue -0.2cm
\caption{\noindent   
Same as Fig.~\ref{fig:mio} but for the solution based 
on NSNI with $d$-quarks in the parameter space of 
$\varepsilon (\equiv\epsilon_d)$ 
and $\varepsilon' (\equiv{\epsilon'}_d)$. 
} 
\label{fig:d-quarks} 
\end{figure}

\begin{figure}[ht] 
\begin{center} 
\epsfig{file=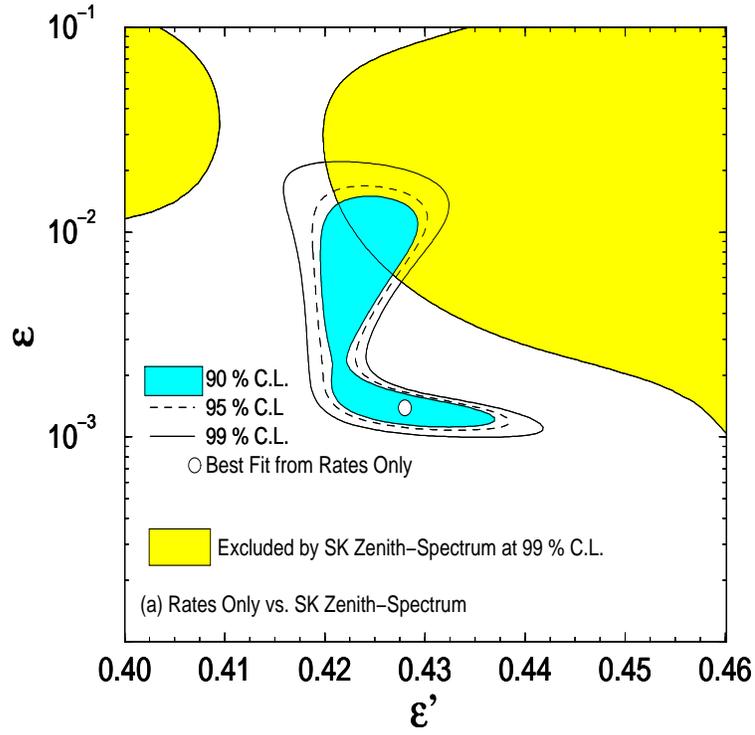,height=10.0cm,width=10.0cm}  
\epsfig{file=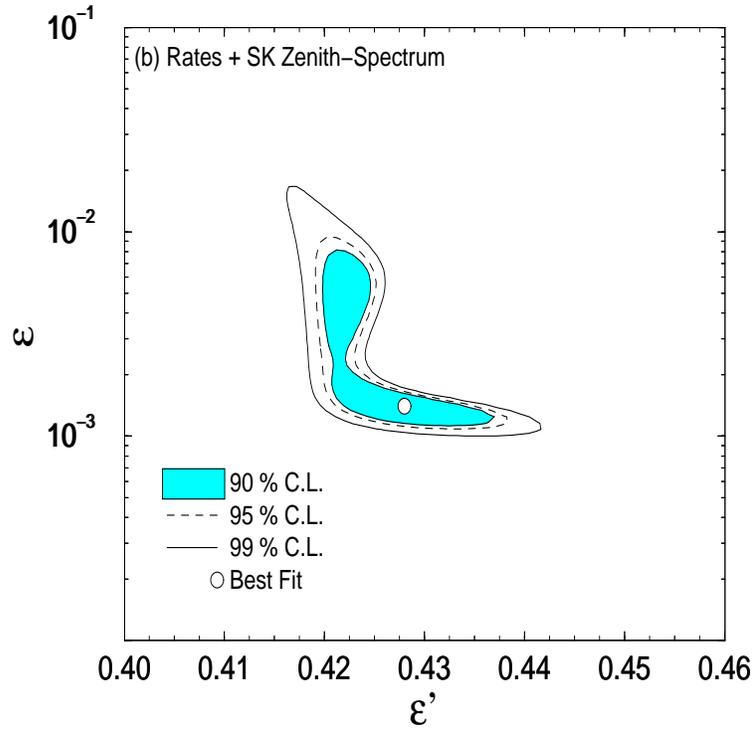,height=10.0cm,width=10.0cm}  
\end{center} 
\vglue -0.2cm
\caption{\noindent   
Same as in Fig.~\ref{fig:d-quarks} but for NSNI with  $u$-quarks
in the parameter space of 
$\varepsilon (\equiv\epsilon_u)$ 
and $\varepsilon' (\equiv{\epsilon'}_u)$. 
} 
\label{fig:u-quarks} 
\end{figure}

\begin{figure}[ht] 
\begin{center} 
\epsfig{file=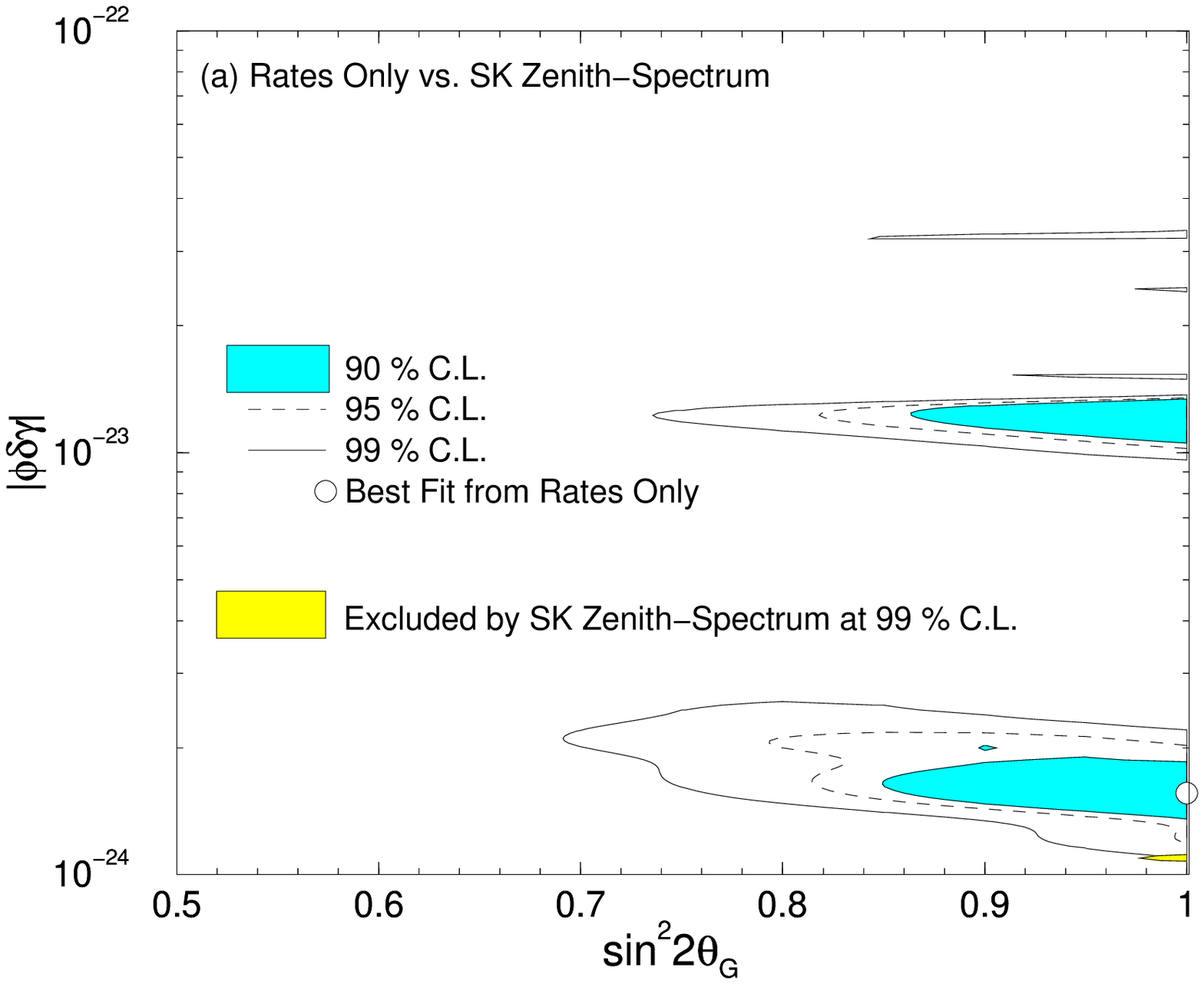,height=10.0cm,width=10.0cm}  
\vglue 0.5cm  
\epsfig{file=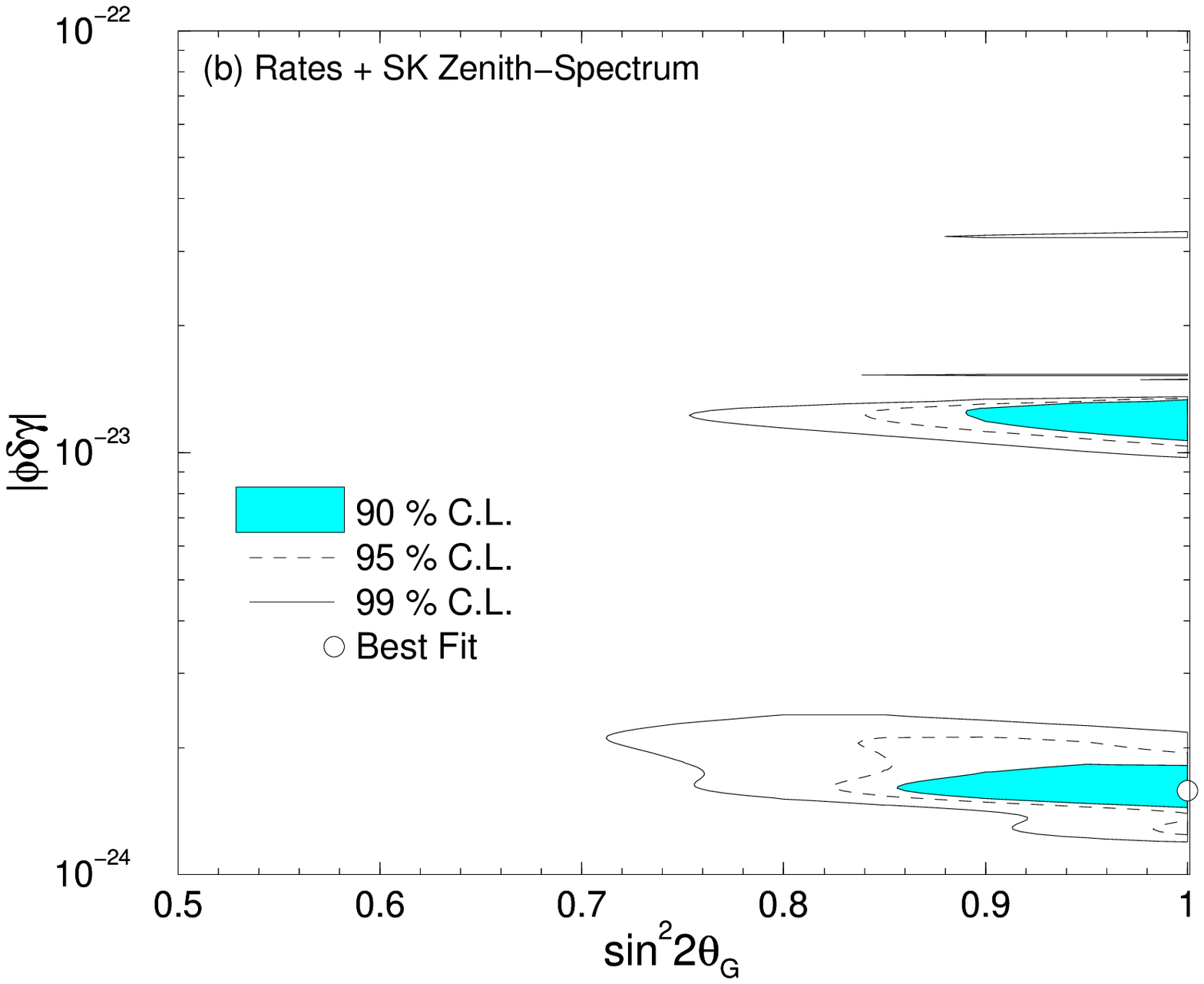,height=10.0cm,width=10.0cm}  
\end{center} 
\caption{\noindent   
Same as Fig.~\ref{fig:mio} but for the VEP solution
in the $\sin^2 2\theta_G-|\phi \delta \gamma |$ plane.
Notice in the upper panel the very small exclusion 
region at 99 \% C.L. from the SK zenith-spectrum 
information appeares at $\sin^2 2\theta_G\sim 1$ and 
$|\phi \delta \gamma |\sim 10^{-24}$.} 
\label{fig:vep} 
\end{figure}

It is worthwhile to note that for the MIO and NSNI cases, 
Figs.\ref{fig:mio}(a), \ref{fig:d-quarks}(a) and \ref{fig:u-quarks}(a), 
in some parameter regions, there is a significant overlap between 
the region allowed by rates only and the one excluded by 
the SK zenith-spectrum information. 
Namely, there is a strong conflict between the  rates and the  
SK zenith-spectrum fit for some parameters and this is 
the reason why the allowed region decreases significantly 
when both data are combined as we can see in Figs. \ref{fig:mio}(b), 
\ref{fig:d-quarks}(b) and \ref{fig:u-quarks}(b). 
On the other hand, for the RSFP and VEP cases, there is no 
such kind of strong conflict between rates and and SK 
zenith-spectrum fit and therefore, the allowed region for 
the combined data is rather similar to the one determined 
only by the rates as we can see in Figs. \ref{fig:rsfp} 
and \ref{fig:vep}. 

\vskip2cm

\section{Discussions}
\label{sec:conc}

Even though the conventional mass induced oscillation mechanism,
which is theoretically well motivated, 
can be considered as the most plausible solution 
to the solar neutrino problem, 
it is important to realize that solutions based on 
New Physics in the neutrino sector, such as 
large neutrino magnetic moment, 
neutrino flavor-changing and non-universal processes, 
violation of the equivalence principle, 
can still be viable ones providing a fit to
the solar neutrino data which is comparable 
to the one based on the conventional 
mass-induced neutrino oscillation, as we showed 
in this work. 

We also note that some of these mechanisms which do not
require neutrino mass in the fit to the solar neutrino data 
do have some close relation with neutrino mass generation. 
For example, in our phenomenological approach, we ignored
neutrino mass in the solution based on non-standard 
neutrino interactions but there is no available 
model that prevents neutrinos to acquire mass 
at radiative level, if flavor changing and flavor
non-universal interactions with quarks are present, 
even if no tree-level mass terms appear in the model.

It is fundamental to mention that it is difficult to 
explain the atmospheric neutrino problem~\cite{AN} as well as 
the LSND anomalies~\cite{LSND} by these alternative 
mechanisms~\cite{failed}. 
Furthermore, while mass induced oscillation and 
resonant spin-flavor precession  solutions to 
the SNP can be easily conciliated with the standard neutrino
oscillation solution to the atmospheric neutrino problem, 
it is not a trivial task to answer if the non-standard 
flavor-changing and non-universal neutrino interactions  
and violation of equivalence principle solutions to 
the SNP modify or even damage this standard solution 
to the atmospheric neutrino anomaly~\cite{hybrid}.  

Having this picture in mind we conclude that the no specific 
solution is preferred by the current solar neutrino data, 
although some solutions may have difficulties in reconciling
atmospheric neutrino observations. 
We emphasize that the solar neutrino observations alone
can not yet conclude if neutrinos have non-vanishing mass 
or magnetic moment. 
Furthermore, no stringent limit on the existence of
NSNI nor on VEP can be currently set by the solar 
neutrino data.

The ultimate goal of the solar neutrino observations is, of 
course, to perform  a direct experimental identification 
of the solution.
For this purpose, we will have to wait for the upcoming solar 
neutrino experiments, which hopefully will provide a lot of new 
informations, to reveal the true nature of the flavor 
conversion mechanism which is behind the SNP. 
For instance, low energy solar neutrino experiments, such as 
Borexino~\cite{borexino}, can possibly discriminate among 
the solutions considered in this work, as can be seen in 
Fig. 7 of Ref.~\cite{nuno-lownu}.

In the near future, a new reactor experiment, KamLAND~\cite{kamland}, may 
measure $\nu_e$-disappearance. For all mechanisms studied here, with
exception of LMA, we expect no significant disappearance in KamLAND
 because the baseline is too short to developed oscillation. 
Then a positive evidence for neutrino oscillation in
KamLAND can establish LMA solution. 
For the negative evidence no
conclusion can be drawn to favor a particular mechanism
studied in this work. A discussion about the consequences of KamLAND
experiment on the  mechanism  of neutrino mass
and mixing hypothesis can be found in Ref.~\cite{gouvea}.

\vskip 0.5cm
Summarizing our conclusions,

\begin{enumerate} 

\item 
The present solar neutrino data only by themselves can not discard, 
{\em a priori}, any of the solutions discussed here:
(a) neutrino mass and mixing, 
(b) non-vanishing neutrino magnetic moment, 
(c) the existence of non-standard flavor-changing 
and non-universal neutrino interactions and 
(d) the tiny violation of the equivalence principle. 
We refer to Tables \ref{tab:rates} and \ref{tab:combined} for comparison.
All solutions have confidence level over $60\%$ C.L. providing 
a very good fit to the solar neutrino data;  

\item 
A very robust statement is that LMA is the best preferred solution
for mass induced oscillation scenario, whereas the SMA is 
the worst one;

\item Future experiments could test these different scenarios and
possibly discard some of them.
\end{enumerate}  
 
\section*{ Acknowledgements}
This work was supported by Funda\c{c}\~ao de
do Estado de S\~ao Paulo (FAPESP) and  by Conselho Nacional
de Ci\^encia e Tecnologia (CNPq). 



\begin{table}[h]   
\caption[Tab]{Observed solar neutrino data used in this
analysis presented together with the theoretical predictions  
of ``BP2000 SSM+New $^8$B''~{\protect\cite{BP00,concha-novo}}.
}
\begin{center}
\begin{tabular}{cccc}
Experiments & Observed Rates & SSM Predictions & Units \\ 
\hline
Homestake & 2.56 $\pm$ 0.23 &  $8.59^{+1.1}_{ -1.2}$ & SNU \\
%
%
SAGE & 77.0 $\pm$ 6. $\pm$ 3 & $130^{+9}_{ -7}$ & SNU \\
GALLEX + GNO & 73.9 $\pm$ 4.7 $\pm$ 4.0  & $130^{+9}_{ -7}$ & SNU \\
SK & 0.391 $\pm$ 0.014 & $1.00^{+0.14}_{-0.15}$  & 
$5.93\times 10^6$ cm$^{-2}$s$^{-1}$ \\
SNO CC & 0.296 $\pm$   0.024 & $1.00^{+0.14}_{-0.15}$  &  
$5.93\times 10^6$ cm$^{-2}$s$^{-1}$  \\ 
\hline 
\end{tabular}
\label{tab1}
\end{center}
\end{table}

\begin{table}[bt]
\caption{Observed SK zenith-spectrum event rates in 
unit of the BP2000~{\protect\cite{BP00}}, taken from 
Table I of Ref.~{\protect\cite{sk_zensp}}.  Errors are only statistical. 
For the energy-uncorrelated and energy-correlated systematic 
uncertainties, see Table I in Ref~\protect\cite{sk_zensp}.}
\newcommand{\plumi}[2]{\matrix{+#1\\[-1.2mm]-#2}}
\begin{center}
\begin{tabular}{l|ccccccc}
\hline
& 
\multicolumn{7}{c}{Observed rates and statistical errors in units of SSM} \cr 
\hline
& 
Day   &  Mantle 1  &  Mantle 2       &  Mantle 3       &
Mantle 4        &  Mantle 5       &  Core           \cr 
\ \ $\cos\theta_z$ 
& 
-1.00--0.00     &  0.00--0.16     &  0.16--0.33     & 0.33--0.50      &
 0.50--0.67     &  0.67--0.84     &  0.84--1.00  \cr 
$E_e$ (MeV) 
\cr
\hline
 5.0--5.5 & \multicolumn{7}{c}{0.436$\pm$0.046} \cr 
 5.5--6.5 & 
0.431$\pm$0.022 & 0.464$\pm$0.060 & 0.410$\pm$0.055 & 0.442$\pm$0.048 &
0.453$\pm$0.048 & 0.495$\pm$0.054 & 0.434$\pm$0.058 \cr 
 6.5--8.0 & 
0.461$\pm$0.013 & 0.524$\pm$0.036 & 0.506$\pm$0.033 & 0.438$\pm$0.028 &
0.466$\pm$0.027 & 0.424$\pm$0.030 & 0.409$\pm$0.033 \cr 
 8.0--9.5 &
0.437$\pm$0.014 & 0.449$\pm$0.038 & 0.482$\pm$0.036 & 0.460$\pm$0.031 &
0.503$\pm$0.031 & 0.461$\pm$0.034 & 0.439$\pm$0.037 \cr
 9.5--11.5 &
0.434$\pm$0.015 & 0.432$\pm$0.042 & 0.493$\pm$0.040 & 0.446$\pm$0.034 &
0.448$\pm$0.034 & 0.435$\pm$0.037 & 0.484$\pm$0.044 \cr
11.5--13.5 & 
0.456$\pm$0.026 & 0.496$\pm$0.071 & 0.290$\pm$0.055 & 0.394$\pm$0.053 &
0.477$\pm$0.056 & 0.439$\pm$0.061 & 0.465$\pm$0.068 \cr
13.5--16.0 &
0.482$\pm$0.056 & 0.532$\pm$0.155 & 0.775$\pm$0.171 & 0.685$\pm$0.141 & 
0.607$\pm$0.130 & 0.471$\pm$0.128 & 0.539$\pm$0.153 \cr
16.0--20.0 & \multicolumn{7}{c}{0.476$\pm$0.149} \cr
\hline
\end{tabular}
\label{tab:spzenith}
\end{center}
\end{table}

\begin{table}[h]   
\caption[Tab]{Comparison of the existing solutions to the SNP
when only the total rates are taken into consideration in 
the statistical analysis. For each one of the indicated mechanisms 
the best fit values of the relevant parameters are shown, 
followed by the corresponding $\chi^2_{\text{min}}$ and its confidence 
level. In the first row we see the result of  the poor fit of the SSM
predictions to the data. All other solutions present 
2 free parameters to fit 5 experimental data points, resulting in 3 d.o.f.. 
}
\begin{center} 
\begin{tabular}{ccccc} 
& & & & \\
Mechanism & & & $\chi^2_{\text{min}}$ & C.L.\\
& & & & \\
\hline
& & & & \\
SSM with no oscillation 
& & & 62.5 (5 d.o.f) & $4\times 10^{-12}$ \\
& & & & \\
\hline 
& & & & \\
MIO & $\delta m^2 ({\rm eV}^2)$ & $\tan^2\theta$&  & \\
& & & & \\
Vacuum &  $6.80\times 10^{-11}$&
           0.425 & 3.02 & 39\%  \\
SMA     &  $7.91\times 10^{-6}$ &
           $1.59 \times 10^{-3}$ & 4.60 & 20\%  \\
LMA     &  $2.80\times 10^{-5}$ &
           0.320  & 2.62 & 45\%  \\
LOW     &  $1.05\times 10^{-7}$ &
           0.743  & 10.5 & 1.5\%  \\
& & & & \\
\hline 
& & & & \\
RSFP     & $\delta m^2 ({\rm eV}^2)$ & $B_{\text{max}} ({\rm kG})$ &  &  
\\ & & & & \\
         & $1.11\times 10^{-8}$ & 338 & 1.01 & 80\%  \\
& & & & \\
\hline
& & & & \\
NSNI & $\epsilon'$ & $\epsilon$   & & \\
& & & & \\
$d$-quarks    & 0.599 & $3.22 \times 10^{-3}$ & 0.72 & 87\% \\

$u$-quarks    & 0.428 & $1.39 \times 10^{-3}$ & 0.73 & 87\% \\
& & & & \\
\hline 
& & & & \\
VEP          & $|\phi\Delta\gamma| $ & $\sin^2 2\theta_G$ & &  \\
& & & & \\
 & $1.56\times 10^{-24}$ & 1.0 & 6.23 & 10\% \\
& & & & 
\end{tabular} 
\end{center} 
\label{tab:rates}   
\end{table}

\begin{table}[h]   
\caption[Tab]{Same as in Table~\ref{tab:rates} but for
the statistical analysis taking into
consideration not only the total rates but also the energy spectrum 
and zenithal dependence of the data.}
\begin{center} 
\begin{tabular}{ccccc} 
& & & & \\
Mechanism & & & $\chi^2_{\text{min}}$ & C.L.\\
& & & & \\
\hline
& & & & \\
SSM with no oscillation & & & 100.0 (48 d.o.f) & $1.6\times 10^{-5}$ \\
& & & & \\
\hline 
& & & & \\
MIO & $\delta m^2 ({\rm eV}^2)$ & $\tan^2\theta$&  & \\
& & & & \\
Vacuum &  $4.65\times 10^{-10}$&
           1.89 & 46.1 & 47\%  \\
SMA     &  $4.93\times 10^{-6}$ &
           $4.35 \times 10^{-4}$ & 61.5 & 6.3\%  \\
LMA     &  $6.15\times 10^{-5}$ &
           0.349  & 38.7 & 75\%  \\
LOW     &  $1.01\times 10^{-7}$ &
           0.783  & 45.0 & 38\%  \\
& & & & \\
\hline 
& & & & \\
RSFP     & $\delta m^2 ({\rm eV}^2)$ & $B_{\text{max}} ({\rm kG})$ &  &  
\\ & & & & \\
         & $1.22\times 10^{-8}$ & 440 & 38.4 & 78\%  \\
& & & & \\
\hline
& & & & \\
NSNI & $\epsilon' $ & $\epsilon$ & & \\
& & & & \\
$d$-quarks    & 0.599 & $3.23 \times 10^{-3}$ & 37.9 & 80\% \\

$u$-quarks    & 0.428 & $1.40 \times 10^{-3}$ & 37.9 & 80\% \\
& & & & \\
\hline 
& & & & \\
VEP          & $|\phi\Delta\gamma| $ & $\sin^2 2\theta_G$ & &  \\
& & & & \\
& $1.59\times 10^{-24}$ & 1.0 & 42.9 & 60\% \\
& & & & 
\end{tabular} 
\end{center} 
\label{tab:combined}   
\end{table} 
\vglue -2cm

%
%
%
%
\vglue -1.2cm  

\newpage

\vglue 0.3cm  
 
\newpage

\vglue 0.3cm  
 
\newpage

\vglue 0.3cm  

\newpage
\vglue 0.6cm

\end{document}